\documentclass{article}
\usepackage{spconf,amsmath,graphicx}
\usepackage{tikz}
\usepackage{pgfplots}
\usepackage{verbatim}
\usepackage{ushort}
\usepackage{url}
\usepackage{hyperref}

 \def\AG#1{\textcolor{red}{}}
\title{Semi-supervised transfer learning for language expansion of end-to-end speech recognition models to low-resource languages}

\name{Jiyeon Kim, Mehul Kumar, Dhananjaya Gowda, Abhinav Garg, Chanwoo Kim}

\address{Speech Processing Lab, AI Center, Samsung Research, South Korea}

\email{\{jstacey7.kim, mehul3.kumar, d.gowda, abhinav.garg, chanw.com\}@samsung.com}

\begin{document}
\ninept
\maketitle
\begin{abstract}
In this paper, we propose a three-stage training methodology to improve the speech recognition accuracy of low-resource languages.
We explore and propose an effective combination of techniques such as transfer learning, encoder freezing, data augmentation using Text-To-Speech (TTS), and Semi-Supervised Learning (SSL).
To improve the accuracy of a low-resource Italian ASR, we leverage a well-trained English model, unlabeled text corpus, and unlabeled audio corpus using transfer learning, TTS augmentation, and SSL respectively. 
In the first stage, we use transfer learning from a well-trained English model. This primarily helps in learning the acoustic information from a resource-rich language. This stage achieves around $ 24\% $ relative Word Error Rate (WER) reduction over the baseline.
In stage two, We utilize unlabeled text data via TTS data-augmentation to incorporate language information into the model. We also explore freezing the acoustic encoder at this stage. TTS data augmentation helps us further reduce the WER by $\sim$ 21 \% relatively.
Finally, In stage three we reduce the WER by another 4\% relative by using SSL from unlabeled audio data.
Overall, our two-pass speech recognition system with a Monotonic Chunkwise Attention (MoChA) in the first pass and a full-attention in the second pass achieves a WER reduction of $\sim$ 42\% relative to the baseline. 
\\

\end{abstract}
\vspace{-2mm}
\section{Introduction}
\label{sec:intro}
\vspace{-2mm}
Recent advances in deep learning techniques have enabled the training of end-to-end (E2E) automatic speech recognition (ASR) \cite{kim2020review} systems consisting of a single neural network model.
The E2E model architecture has a simpler training pipeline and better modeling capabilities compared to conventional architectures such as DNN-HMM systems. 
With these advantages, E2E speech models are widely used by state-of-the-art ASR systems for server-side and on-device applications~\cite{li2020fast,kim2019endtoend,2pass, garg2020hierarchical}. 
However, training E2E models from scratch for a new language requires a lot of data to achieve high accuracy \cite{lugosch2019speech, garg2020streaming}. 
Obtaining valid and reliable transcriptions for speech data is costly, time-consuming, and refining both text and speech data is laborious work.

To overcome the requirement of a large amount of labeled data, there have been several efforts to train E2E models with a smaller amount of data. 
For example, ~\cite{nguyen-chiang-2017-transfer, DBLP:journals/corr/KunzeKKKJS17, miao_y, BESACIER201485} explored transfer learning from a resource-rich language to a low-resource language. 
It has been also shown that using other language corpus and multilingual training, the performance for a low-resource language can be improved~\cite{matsuura2020generative}. 
In other related works, \cite{DBLP:journals/corr/abs-1711-02207} suggests a universal character set and creates a language-specific gating mechanism to increase the network's modeling power.
In \cite{DBLP:journals/corr/abs-1904-05862}, unsupervised pretraining is used to improve acoustic model training, and semi-supervised learning was used in \cite{Nallasamy2012SemisupervisedLF} to leverage the lack of sufficient amounts of labeled data. 
Recent works like wav2vec 2.0 \cite{baevski2020wav2vec} particularly rely on self-supervision to train a task-agnostic encoder from the huge amount of unlabeled data before applying a task-oriented supervised training with limited data.\\
\indent In this paper, we explore methods to improve the performance of an ASR model for an under-resourced language with the limited amount of labeled data. We split the training process into three stages, each using a different training paradigm to leverage either a model trained on resource-rich language or unlabeled text data or unlabeled audio data.\\
\indent During each training stage, we use different training methodologies such as transfer learning from a resource-rich language, spectral augmentation, Text-To-Speech (TTS) augmentation, encoder freezing, and beam score filtering based semi-supervised learning. 
For our experiments, we use one of our recently proposed two-pass streaming architecture described in ~\cite{2pass} with Monotonic Chunkwise Attention (MoChA) decoder in the first pass and a Bidirectional-encoder Full-Attention (BFA) decoder in the second pass.\\
\indent To the best of our knowledge, this is one of the first works that explore leveraging a resource-rich language, unlabeled text data, and unlabeled audio data using transfer learning,  TTS data augmentation, and semi-supervised techniques respectively, for training a streaming end-to-end MoChA attention-based model for a low-resource language.\\
\indent As a case study, we use English and Italian as our high and low-resource languages.
The English corpus has around 11K hours of transcribed data.
The Italian corpus with around 4000 hours of labeled data is divided into two more subsets of around 400 hours and 40 hours each. 
The 40 hours subset represents the low-resource labeled speech corpus.\\
\indent We show that there exists a huge difference in Word Error Rates (WER) when models are trained with different amounts of data from scratch using standard training practices used for E2E ASR models. 
We design a training methodology including a beam score filtering-based semi-supervised training method.\\
\indent We show that transfer learning using a well-trained model from a resource-rich language helps achieve better accuracy compared to training models from scratch in a new language, especially when the amount of labeled data available is limited.
TTS-generated audios are typical of sub-optimal quality as compared to real-world audios.
To mitigate this effects of 
TTS audio, we propose freezing of encoder when applying TTS data augmentation. Using the proposed training methodology we reduce the performance gap of the model trained with 40 hours as compared to 400 hours and 4000 hours models by around 73\% and 62\%, respectively.

\vspace{-2mm}
\section{Two-pass MoChA-BLSTM streaming model architecture}
\label{sec:format}
\begin{figure}[t]
\centering
\includegraphics[width=3.5cm,height=4cm,trim={0mm 2mm 0mm 2mm},clip]{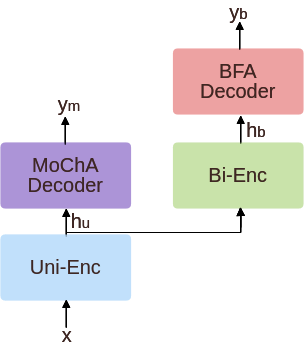} 
\vspace{-2mm}
\caption{\label{fig:2pass} The structure of a two-pass MoChA-BLSTM full attention end-to-end streaming speech recognition system. 
}
\vspace{-5mm}
\end{figure}

\vspace{-2mm}
In this paper, we use a two-pass streaming end-to-end ASR architecture similar to the one described in ~\cite{2pass}.
To achieve high speech recognition accuracy and streaming capabilities simultaneously, we employ a two-pass MoChA-BLSTM model shown in Fig. \ref{fig:2pass}.
The two-pass model used in this paper consists of a shared encoder stack (Uni-Enc) of six layers of uni-directional Long Short-Term Memory (LSTM)  with intermediate maxpool layers to reduce the overall temporal rate. 
A MoChA-based decoder (MoChA Decoder) is attached to the shared encoder to provide streaming capabilities. 
An additional encoder stack of a bi-directional LSTM layer (Bi-Enc) is placed on top of the Uni-Enc encoder.
To reduce the total number of model parameters and keep the model footprint small, Bi-Enc contains just a single backward LSTM.
The output of Bi-Enc is fed to a full-attention decoder and is referred to as the bidirectional-encoder full-attention decoder (BFA Decoder).
The streaming MoChA decoder generates the intermediate speech recognition result. At the end of the utterance, the BFA decoder generates the final speech recognition result with a better WER.

\vspace{-3mm}
\section{Training strategies for low-resource languages}
\label{sec:train_strategies}
\vspace{-3mm}
In this paper, we focus on building an ASR model for low-resource language with a limited amount of transcribed speech data.  
We aim to reduce the gap in performance between this model and a model trained with 10 to 100 times more data.
To achieve this, we apply a multi-stage training strategy with different techniques used during each training stage. 
We split the training into 3 stages.
In the first stage, we use transfer learning by borrowing pre-trained weights from a well-trained model of a resource-rich language. The borrowed model parameters are updated using the limited amount of labeled data available for the target language.
The primary objective of this stage is to transfer the acoustic information from a high-resource to a low-resource language.
In the second stage, we use a Text-To-Speech (TTS) based data augmentation to leverage the more readily available text-only data for any low-resource language. 
We show that freezing the encoder during this stage is an important step so that the acoustic information learned from the real speech in stage one is not disturbed adversely using TTS data which can be possible, noisy, and low-quality.
This stage only incorporates additional linguistic information into the model.
In the final step, we use Semi-Supervised Learning (SSL) based approach to using unlabeled audio-only data that may be available for the target language. 
%
\vspace{-1mm}
\subsection{Stage One: Transfer Learning}
\vspace{-1mm}
\label{sssec:subhead}
Transfer learning can be of two types, one that involves the transfer of soft knowledge, and the other that involves the transfer of hard knowledge~\cite{nguyen-chiang-2017-transfer, DBLP:journals/corr/KunzeKKKJS17, crosslanguagetransfer}.
Transfer of soft knowledge refers to teacher-student learning strategies that minimizing a cost function to reduce the output mismatch between teacher and the student.
This strategy is more suited for applications like model compression or moving from a mono-lingual ASR to a bi-lingual or multi-lingual ASR where one may want to retain the performance of the teacher.
Transfer of hard knowledge refers to borrowing the pretrained weights from a teacher model into a student model before continuing with independent training of the student.
\AG{THERE is some problem with this statement but not sure what} In this paper, we use hard transfer learning as a first step since we train the model independently for low-resource language by borrowing weights from a previously well-trained resource-rich language.

As the pool of acoustic sounds is universal with significant overlap across different language, we borrow the pretrained model weights from a high-resource language to a low-resource language.

However, the exact realization of these phonetic units may vary based on the regional or linguistic traits and may need some fine-tuning to the sounds of a target language.
Also, the phonetic and linguistic grammar (combination or sequence of phones and words) are different between languages.
Hence the encoders which encode the acoustic information in the audio signal should be more similar across languages as compared to decoders.
To verify this hypothesis, we try to freeze the borrowed encoder before continuing with supervised training from the limited labeled data available for the target language.
\vspace{-1mm}
\subsection{Stage Two: Text-to-Speech data augmentation and Encoder freezing}
 \label{sec:tts_enc_freeze}
\vspace{-1mm}
Text data is more readily available for most languages as compared to audio data or transcribed audio-text pair. 
And can be easily obtained in large quantity using web-crawling, digital books, etc.
Due to the abundance of text data, many methods have been explored to leverage it for building ASR models.
First, an external language model can be built and used along with the ASR model using shallow fusion during inference \cite{toshniwal2018comparison}.
Second, borrowing the weights of a pretrained language model to initialize the language modeling components of end-to-end models.
This can be readily done for a recurrent neural network transducer (RNNT) architecture with a separate predictor block but would require a more careful integration into an Attention-based Encoder-Decoder (AED) model.
The other option is to use a TTS engine for the target language to synthesize audio data from the text to train the ASR model.
A variant to this approach is to use a Text-To-Encoder (TTE) approach wherein a parallel stack of the text-only encoder is used along with the standard encoder in an ASR that accepts audio-only inputs~\cite{baskar2019tte}.
During the training stage, the text corresponding to the audio data is passed through the text-only encoder to produce a hidden text-encoder embedding that is forced to match the audio-encoder embeddings by minimizing the Kullback-Liebler divergence between the two.
In this paper, we propose to use the TTS based data augmentation as it generally shows better performance compared to TTE based approach~\cite{baskar2019tte} except for the need of a TTS engine for the target low-resourced language.
The Google Speech python library \cite{google-speech} can be readily used to read text using the Google Translate TTS APIs.
%
\\
\indent It has been observed that data augmentation with TTS performs better than the TTE approach~\cite{baskar2019tte}.
However, this begs the question as to whether the single speaker, possible monotonous,  and poor audio-quality TTS signals are good to update the encoder weights that learn to encode the acoustic information.
Some of these limitations can be mitigated to a certain extent with the use of different perturbation techniques like speed, tempo, pitch, vocal tract length~\cite{C_Kim_ASRU_2009_1}, and data augmentation techniques like spectral augmentation, Room Impulse Response (RIR), and noise-based acoustic environment simulators~\cite{C_Kim_ASRU_2009_2}.
Another way to handle this would be to freeze the audio encoder and allow the update or learning of attention and decoder parameters of the E2E model.
To reduce the adverse effects of acoustic features from TTS synthesized data, we freeze the encoder while retraining the model with TTS data \cite{encoderfreeze}. 
We show that encoder freezing during TTS data augmentation improves model accuracy. 
%
%
\vspace{-2mm}
\subsection{Stage Three: Filtering ASR hypotheses through Semi-supervised learning}
\vspace{-1mm}
Semi-Supervised Learning (SSL) based training is widely used for low-resource data \cite{semi-super,gupta2018semi, singh2020large, nst2020}, as well as learning representations from large speech corpus without corresponding labeled transcriptions~\cite{baevski2020wav2vec}.
In this step, the current ASR model is used to generate text hypotheses for the unlabeled audio data which can now be used to train the same ASR model.
To generate a good quality model output using semi-supervised learning, we first improve the model with different algorithms to reduce the WER as much as possible. We apply semi-supervised learning during each training stage, and figure out semi-supervised training works when the base model performance is relatively good. 
To verify the efficacy of SSL on models of different accuracies, we apply SSL both before and after applying TTS data augmentation.
It is also widely known that even with a large unlabeled audio dataset, the performance of the SSL trained model degrades when all the data is used indiscriminately \cite{nst2020}. 
To further improve the model's performance using semi-supervised training, we apply a filtered beam search-based semi-supervised training in stage 3.
%

We propose to use the beam posterior probability score of the top hypothesis to filter the data.
Other more sophisticated iterative filtering methods can be used which try to balance the data across different lengths of audio data~\cite{nst2020}.
Also, using a well-trained language model with a large text corpus will help in generating good quality ASR hypotheses.
However, in this paper, we use a simple single step beam score based filtering without any external language model.
%
\vspace{-2mm}
\section{Experiments}
\vspace{-2mm}
\label{sec:page}
\subsection{Datasets}
\vspace{-1mm}
\label{sssec:subhead}
The experiments in this paper are carried out on anonymized internal datasets for the English and Italian languages collected over several mobiles and visual display devices.
The details of the datasets are given in Table~\ref{tab:data}.
The English dataset consists of around 11K hours of transcribed data, which is further augmented to generate an equal amount of noisy and RIR simulated data. The Italian data ITA4kh consists of around 4k hours of transcribed data.
The Italian data is split into two more subsets, ITA400h and ITA40h with 400 and 40 hours, respectively.
The subset ITA40h is used to simulate a low-resource language setting.
The text and audio data in ITA400h after removing ITA40h are used separately for TTS data augmentation as well as SSL experiments.
Around 20K and 2K utterances are held out from the randomized full Italian corpus as validation and test sets respectively.
\begin{table}[t]
\caption{\label{tab:data} Details of the English and Italian datasets.}
\vspace{1mm} 
\centering
\begin{tabular}{|c|c|c|c|c|}
\hline
Dataset & ENG & ITA4kh & ITA400h & ITA40h  
\\\hline\hline
No. of Utts & $\sim$9M & $\sim$6.6M & $\sim$660K & $\sim$66K\\\hline
Duration (hr) & 11,014 & 4,247 & 437 & 44 \\\hline
\end{tabular}
\vspace{-5mm}
 \end{table}







\vspace{-2mm}
\subsection{Experimental Setup}
\vspace{-1mm}
\label{sssec:subhead}
For all the experiments, the feature extraction is based on a power-mel filterbank with a power coefficient of ($\cdot)^{1/15}$ \cite{Kim2019} to create a 40-dimensional feature vector. 
We use a window size of 25$ms$ with an interval of 10$ms$ between successive windows. 
The text transcriptions are used to generate Byte Pair Encodings (BPE) labels with a vocab size of around 10K for both English and Italian. We enable a dropout rate of 30\% for the encoder layer, and 10\% of label smoothing for the output probability. During the inference, we use a beam size of 12 as default. All the training and testing data is 16 kHz audio. The shared encoder Uni-Enc has six LSTM layers with 1536 cells each. Bi-Enc consists of a single backward LSTM later with 1536 cells.
The decoder BFA Decoder and MoChA Decoder consist of a single LSTM layer of 1000 cells each.
Spectral augmentation is enabled throughout all stages of training by default.
To achieve faster and better alignments between the input and output sequences, we train the model with a joint loss \cite{suyounctc} function with two Connectionist Temporal Classification (CTC) losses for the two encoders (Uni-Enc and Bi-Enc) and two Cross-Entropy (CE) losses for the two decoders (MoChA and BFA) given by,
\begin{align}
L_{total}  = L^{Uni-Enc}_{ctc} +  L^{Bi-Enc}_{ctc} + L^{MoChA}_{ce}  + L^{BFA}_{ce}.
\end{align}






\vspace{-4mm}
\subsection{Results}
\vspace{-1mm}
\label{sec:page}
To start with, we set up baseline E2E ASR models for Italian using different amounts of data from scratch.
The WERs of different models are shown in Table~\ref{tab:ita_base}.
The ITA400h baseline model is around 26\% relatively poorer than the ITA4Kh model, while the ITA40h baseline is almost 58\% and 69\% poorer than the 400h and 4Kh models respectively.
It can be seen from the results that the performance of the ASR models degrades drastically with lesser amounts of data.
Our efforts are primarily to bridge the gap in performance between the ITA40h model as compared to the other two data rich models.
To see the impact of transfer learning when different amounts of supervised data are at hand, we borrow weights from an English model trained using the 11K hours corpus.

\begin{table}[t]
\centering
\caption{\label{tab:ita_base} Performance of models in terms of WERs (\%) after Transfer Learning (TL) from a pretrained English model to Italian using different amounts of data, as compared to Italian baseline models trained from scratch.}
\vspace{1mm}
\begin{tabular}{|c|c|c|c|}
\hline
Model                   & ITA4kh   & ITA400h  & ITA40h \\\hline\hline
Ita Baselines (scratch) & 11.65 & 15.79 & 37.56\\\hline
TL from Eng to Ita & 10.78 & 15.33 & 28.50 \\\hline
\end{tabular}
\vspace{-5mm}
\end{table}
\vspace{-1mm}
\subsubsection{Effect of transfer learning}
It can be seen from Table~\ref{tab:ita_base}, that transfer learning outperforms the models trained from scratch by around 7.5\%, 3\%, and 24\% relative for the 4Kh, 400h, and 40h scenarios, respectively.
Notwithstanding a dip in relative improvement for the 400h case, the impact of transfer learning is along expected lines, which is more when the amount of supervised data at hand is less.
The performance of the baseline English model on the Italian test set and the effect of transfer learning by freezing the encoder and only allowing the decoder to train on supervised data is shown in Table~\ref{tab:tl}.

\AG{rewrite this part}
It is known that transfer learning gives good results when the languages are closely related. English and Italian are closely related phonetically but differs in linguistic grammar. Hence, we first try to finetune only the decoder while keeping the encoder frozen. However, we observe that finetuning both encoder as well as decoder gives better performance as it allows both acoustic as well as linguistic fine-tuning.
\begin{table}[t]
\caption{\label{tab:tl} Effect of using transfer learning from a well trained English model and supervised training with around 40 hours of data.}
\vspace{-3mm}
\begin {center}
\begin{tabular}{|l|c|c|}
\hline
Model & ITA400h & ITA40h \\\hline\hline
B1: Ita Baselines & 15.79 & 37.56 \\
B0: Eng Model & - & 107.8 \\\hline
M1: B0 + TL w/ Enc freezing & - & 30.69 \\
M2: B0 + Transfer learning & - & {\bf 28.50} \\\hline
\end{tabular}
\vspace{-6mm}
\end {center}
\end{table}

\vspace{-2mm}
\subsubsection{Effect of TTS data augmentation}
\vspace{-1mm}
In stage two, we study the effect of TTS data augmentation for improving ASR performance.
The impact of TTS data augmentation on models at different accuracy levels is shown in Table~\ref{tab:ttsaug}.
By comparing the WERs of M3 and M4 we can easily conclude that using TTS augmentation provides larger improvements for a better model B2 as compared to B1.
The improvements of model M3 and M4 over their baselines B1 and B2 are 3.26\% and 5.0\% absolute, or 8.6\% and 17.5\% relative, respectively.
This shows that using transfer learning in stage one enhances the gains obtained from TTS augmentation.

As mentioned in Sec~\ref{sec:tts_enc_freeze}, the TTS audio may not be ideal for learning or updating the encoder weights which predominately captures and encodes the acoustic information in a speech signal.
Too much TTS audio may bias or overfit the model to TTS data, with a possible degradation in model performance.
This can be seen from the experiments in Table~\ref{tab:ttsaug}, where the model M5 with a frozen encoder performs better than the M4 without encoder freeze by around 4.4\% relative. 
Also, M5 gives an overall 21.2\% relative improvement over the baseline model B2 after transfer learning, as against the 17.5\% relative improvement obtained without freezing the encoder.

\begin{table}[t]
\caption{\label{tab:ttsaug} Effect of TTS based text data augmentation using around 360h of synthesized audio, with and without encoder freezing. The performance of different models is given in WERs (\%).}
\vspace{-2mm}
\begin{center}
\begin{tabular}{|l|c|c|}
\hline
Model & ITA400h & ITA40h 
\\\hline\hline
B1: Ita Baselines & 15.79 & 37.56 \\
B2 (M2): Transfer learning & 15.33 & 28.50 \\\hline
M3: B1 + TTS Aug  & - & 34.30 \\
M4: B2 + TTS Aug   & - & 23.50 \\ 
M5: B2 + TTS Aug w/ Enc freezing  & - & {\bf 22.45} \\\hline 
\end{tabular}
\end{center}
\vspace{-6mm}
\end{table}
\begin{table}[t]
\caption{\label{tab:ssl} Effect of semi-supervised learning (in \% WERs) in utilizing unlabeled audio data for language transfer.}
\vspace{1mm}
 \resizebox{\linewidth}{!}{
\begin{tabular}{|l|c|c|}
\hline
Model & ITA400h & ITA40h  \\\hline\hline
B2 (M2): Transfer learning  & 15.33 & 28.50 \\
B3 (M5):       TL + TTS Aug w/ Enc freezing & - & 22.45 \\\hline
M6: B2 + SSL w/o filtering & - & 29.08 \\
M7: B3 + SSL w/o filtering & - & 22.72 \\
M8: B3 + SSL w/ Beam Score Filter & - &  {\bf 21.52} \\\hline
M9: B3 + SSL w/ Oracle Ground Truth Filter & - & 20.44 \\\hline
\end{tabular}
}
 \vspace{-4mm}
 \end{table}
\begin{table}[tbh]
\caption{\label{tab:sslfilt} Filtering ASR hypotheses of unlabeled audio data using beam scores (BS) during semi-supervised training. A comparison of WERs (in \%) at different thresholds.}
\vspace{1mm}
  \resizebox{\linewidth}{!}{
\begin{tabular}{|c|c|c|c|c|c|}
\hline
BS filter threshold & ${<}$-1& ${<}$-5 & ${<}$-10 & ${<}$ -50 & all\\\hline\hline
No. of Utts  & 169K & 422K & 548K & 618K & 619K \\\hline
SSL with BS filter & {\bf 21.52} & 21.72 & 21.87 & 22.28 & 22.72 \\\hline
\end{tabular}
}
\vspace{-4mm}
\end{table}

\vspace{-2mm}
\subsubsection{Effect of semi-supervised learning}
\vspace{-1mm}
As the final stage of training for language transfer for ASR models, we explore the use of the SSL from unlabeled audio data.
While several methods have been proposed in the literature, we use the most straight-forward approach of generating text labels for a corpus of unlabeled audio data using our best ASR model so far.
It is well known that better-trained models tend to benefit more from audio-based SSL techniques as they can produce better hypotheses for the unlabeled data.
Results for our experiments with SSL techniques can be seen in Table ~\ref{tab:ssl}.

WERs for models M6 and M7 show that performance can degrade when all of the ASR-generated text labels are used indiscriminately for SSL.
The adverse effect is more when the baseline accuracy of the ASR model is poorer. Comparing B3 and M7 shows that TTS augmentation as stage 2 is adequate as the WER increases in M7 without beam score-based filtering. 
Using a simple ASR hypothesis filtering by using the log posterior probability score of the top beam, it can be seen that the model improves by around 4.1\% relative at stage 3.

Using a language model for generating text labels, different scores for filtering and iterative methods for SSL can help improve the gains further.
To see the maximum gain possible with SSL if we were to use a better metric or score for filtering the hypotheses, we use the oracle ground truth labels to identify all correctly hypothesized data which amounts to around 60\% of the total $\sim$619K utterances (around 400 hours) available in the SSL unlabeled dataset.
Using this oracle ground truth filtered data gives an improvement of around 8.9\% relative as compared to around 4.1\% relative when using a simple automated filtering mechanism.
The performance of filtering by beam scores at different thresholds for filtering is shown in Table~\ref{tab:sslfilt}.
It can be seen that with a harsher threshold we can expect lesser but more reliable transcripts and hence there is a progressive improvement in the gains achieved by SSL.

%
\vspace{-4mm}
  \section{Conclusions}
  \label{sec:con}
  \vspace{-2mm}
  In this paper, we present a sequence of training strategies to develop and improve ASR models for a resource-constrained language. 
  The training involved three stages: transfer learning from an ASR model in a resource-rich language, TTS text data augmentation with encoder freezing, and semi-supervised learning by filtering the automatically generated hypotheses for unlabeled data using beam scores.
  The proposed methodology improved the performance of the 40h model by around 42\% relative to a model trained from scratch using the 40h data.
  It was seen that overall the proposed methodology reduces the gap in performance between the 40h and 400h models by around 73\%, and for 40h and 4Kh models by around 62\%.
  All the performances discussed so far in the paper are for the second pass BFA decoder.
  The first pass MoChA decoder of the final model (M8) gives a WER of 24.84\%, an improvement by 43.9 \% relative, over the 40h baseline model (B1) with a WER of 44.35\%.

\bibliographystyle{IEEEbib}
\bibliography{strings,refs,jiyeon,mybib}

\end{document}